\input phyzzx
\input epsf

\hsize=6.5truein
\vsize=8.5truein

\def\pr#1#2#3#4{Phys.~Rev.~D{\bf #1}, #2 (19#3#4)}
\def\np#1#2#3#4{Nucl.~Phys.~{\bf B#1}, #2 (19#3#4)}
\def\pl#1#2#3#4{Phys.~Lett.~{\bf #1B}, #2 (19#3#4)}

\def\uex{U_{\rm ex}}
\def\vex{V_{\rm ex}}
\def\lag{{\cal L}}                      

\def\phit{{\widetilde\varphi}}          
\def\msbar{{$\rm\overline{MS}$}}        
\def\mubar{{{\bar\mu}^2}}               %
\def\at#1#2{{\left.{#1}\right|_{#2}}}   
\def\attree#1{{\left.{#1}\right|_{J_0, \phi^2_0}}}

\def\wtilde{{\widetilde W}}
\def\gtilde{{\widetilde\Gamma}}

\def\ytilde{{\widetilde Y}}

\def\dk{{\int{d^dk\over(2\pi)^d}\ }}    

\def\phir{{[\varphi_a\varphi_a]_{\rm r}}}
\def\phis{{[\varphi_a\varphi_a]_{\rm s}}}
\def\rxi{{$R_\xi$-gauge}}

%
%
%
%

%
%

\REF\dolanjackiwa{L. Dolan and R. Jackiw, \pr9{2904}74.}

\REF\nielsen{N. K. Nielsen, \np{101}{173}75.}

\REF\buchmuller{W. Buchm\"uller, Z. Fodor and A. Hebecker,
\pl{331}{131}94.}

\REF\prework{For some previous work, see K. Cahill, \pr{52}{4704}95;
K. Okumura, hep-th/9404102; and S. Yokojima, \pr{51}{2996}95.}

\REF\cornwall{J. M. Cornwall, R. Jackiw and E. Tomboulis,
\pr{10}{2428}74.}

\REF\jackiw{R. Jackiw, \pr9{1686}74.}

\REF\collinspaper{J. C. Collins, \np{80}{341}74.}

\REF\collins{J. Collins, {\it Renormalization}, Cambridge University
Press, Cambridge, 1984.}

%
%
%
%

\nopagenumbers
\line{\hfill CU--TP--740}
\line{\hfill hep-ph/9602435}
\vskip 0.5in
\title{An Effective Potential for Composite Operators}
\author{Yue Hu}
\address{Department of Physics \break
        Columbia University \break
        New York, NY 10027}

\vskip 1in
\midinsert
\narrower
\centerline{ABSTRACT}
We study the effective potential for composite operators. Introducing
a source coupled to the composite operator, we define the effective
potential by a Legendre transformation. We find that in three or fewer
dimensions, one can use the conventionally defined renormalized
operator to couple to the source. However, in four dimensions, the
effective potential for the conventional renormalized composite
operator is divergent. We overcome this difficulty by adding
additional counterterms to the operator and adjusting these order by
order in perturbation theory. These counterterms are found to be
non-polynomial. We find that, because of the extra counterterms, the
composite effective potential is gauge dependent. We display this
gauge-dependence explicitly at two-loop order.

\endinsert

\vfill
\endpage

\pagenumbers
\pagenumber=1

\chapter{Introduction}

It is well known that the effective potential for elementary fields is
gauge dependent[\dolanjackiwa]. The effective potential can be used in
the studies of spontaneous symmetry break, inflationary cosmology and
many other problems. It is important to examine if
the gauge dependence of the effective potential causes the physical
quantities to be gauge dependent. Nielsen discovered an important
identity on the gauge-dependence of the effective
potential[\nielsen]. With the Nielsen identity and its variations,
many physical quantities can be proved to be gauge independent.

The gauge dependence of the effective potential arises because the
elementary fields are not invariant under gauge transformation. This
suggests that one might obtain an explicitly gauge-independent result
by defining an effective potential for a gauge-invariant composite
operator[\buchmuller]. We will examine this issue in this paper.
To find the effective potential $U(\sigma)$ for a composite operator
$\varphi^2$[\prework], one introduces a source coupled to this
operator
$$ \lag\rightarrow\lag-J\varphi^2. \eqno\eq $$
The effective potential $U(\sigma)$ is the Legendre transform of the
ground state energy density $w(J)$ with constant external source $J$,
where $\sigma$ is conjugate to $J$. Since the unrenormalized composite
operator $\varphi^2$ is divergent in general, one has to renormalize
it by adding appropriate counterterms to the unrenormalized
operator. In this paper, we will use the following method to calculate
the effective potential for the renormalized operator $[\varphi^2]$:
We will couple the system to two external sources
$$ \lag\rightarrow\lag-J[\varphi^2]-h\varphi=\lag(J)-h\varphi.
	\eqno\eq $$
If we treat the first external source as part of the Lagrangian, and
Legendre transform with respect to the other source $h$, we will get
the effective potential for the modified Lagrangian,
$V_{\lag(J)}(\phi)$. We introduce an intermediate object
$$ Y(\sigma, J, \phi) = J\sigma+V_{\lag(J)}(\phi). \eqno\eq $$
We will show that by minimizing $Y(\sigma, J, \phi)$ with respect to
$J$ and $\phi$, one can get the effective potential $U(\sigma)$ from
the function $Y(\sigma, J, \phi)$.
In three and fewer dimensions, if we use the renormalized
operator $[\varphi^2]$, the effective potential $U(\sigma)$ is a
well-defined finite object. However, in
four dimensions, there does not exit a finite effective potential for
the conventionally defined composite operator $[\varphi^2]$. This is
because in four dimensions, matrix elements with more than one
insertion of $[\varphi^2]$ is divergent in general even when we have
added appropriate counterterms to the composite operator to make
matrix elements with one insertion finite. Therefore, adding a source
of $[\varphi^2]$ causes vacuum energy divergence in the composite
effective potential. We define a finite composite effective potential
by adding extra counterterms to $[\varphi^2]$ and adjusting these
counterterms order by order. We find that at two-loop order, the
counterterms of the modified composite operator become
non-polynomial.

Although the ordinary effective potential is gauge dependent, its
minimal value is gauge independent because of the Nielsen
identity[\nielsen]. Therefore, the composite effective potential is
gauge independent because it is the minimal value of $Y(\sigma, J,
\phi)$ with respect to $J$ and $\phi$. In four dimensions, the extra
counterterms of the modified composite operator invalidate this
argument. If we adopt a minimal subtraction scheme for the
counterterms, we find that the composite effective potential is
explicitly gauge dependent at two-loop order. However, one may add
finite counterterms to the composite operator to make the composite
effective potential gauge independent, but there is no preferred
description to choose finite parts over the other descriptions. Hence
although we can make the composite effective potential gauge
independent, there is no clear description in how to resolve the
arbitrariness.

In Section~{\Roman 2}, we study the physical meanings of the effective
potential and the calculation method that we will use later. In
Section~{\Roman 3}, we illustrate our method with a three-dimensional
example. In Section~{\Roman 4}, we show the need to add extra
counterterms in four dimensions and that the extra counterterms are
non-polynomial by studying an ungauged $O(N)$ model. In
Section~{\Roman 5}, we use the same method to study the effective
potential $U(\sigma)$ for scalar QED and its gauge dependence.

\chapter{Effective Potential and Calculation Method}

Let us consider a quantum field theory in Euclidean space at zero
temperature. For simplicity, we will restrict ourselves to the case
where there is only one elementary field. The generalization to
multiple fields is straightforward. Introducing an external source
$J(x)$ to the elementary field $\varphi(x)$, we can define a
functional of $J(x)$
$$ W[J] =\ln\int\left[{\cal D}\varphi\right]e^{-S+\int d^dx J(x)
	\varphi(x)} \eqno\eq $$
where $S$ is the Euclidean action of the theory and $d$ is the number
of space-time dimensions. We define a new variable $\phi(x)$ by
$$ \phi(x)={\delta W[J]\over\delta J(x)}. \eqno\eq $$
The effective action of the theory is defined as the Legendre
transformation of the functional $W[J]$
$$ \Gamma[\phi] = \int d^dx\ J(x)\phi(x) - W[J]. \eqno\eq $$
We can expand the effective action as a power series of external
momenta. In position space, such an expansion can be written as
$$ \Gamma[\phi] = \int d^dx \left[ V(\phi) + {1\over2}(\partial\phi)^2
	\Gamma^{(1)}(\phi)+\cdots\right]. \eqno\eq $$
The function $V(\phi)$ in the first term of this expansion is called
the effective potential for the elementary field $\varphi$.

Suppose $H$ is the Hamiltonian of the system. Let us ask the following
question: among all states that satisfy the constraint
$$ \bra\psi\varphi(x)\ket\psi=\phi, \eqn\constraint $$
which state has the minimal value of $\bra\psi H\ket\psi$ and what is
the minimal value? Using the method of Lagrange multiplier, one finds
that the minimal value is related to the effective potential by
$$ V(\phi)=J\phi-{W\over\beta\Omega}={{\vev H}_{\rm min}\over\Omega}.
	\eqno\eq $$
Thus we conclude that, at zero temperature, if we require that a
homogeneous system satisfy the condition $\vev{\varphi(x)}=\phi$, the
minimal energy density of the system is the effective potential
$V(\phi)$.

To calculate the effective potential $V(\phi)$, let us shift the field
and define a new field $\phit$ by
$$ \varphi(x)=\phit(x)+\phi. \eqno\eq $$
Next we rewrite the Lagrangian in terms of the new field $\phit(x)$ and
separate out the terms linear in $\phit$:
$$ \lag = \lag'+a_1\phit(x). \eqno\eq $$
By the definition of the effective potential, we have
$$ e^{-\int d^dx V(\phi)} = e^{W-\int d^dxJ\phi}
	= \int\left[{\cal D}\varphi\right]e^{-\int\lag+\int J\varphi
	-\int J\phi}. \eqno\eq $$
In terms of the new variable $\phit(x)$, this equation can be written
as
$$ e^{-\int V(\phi)}=\int\left[{\cal D}\phit\right]
	e^{-\int\lag'+\int(J-a_1)\phit}=e^{W_{\lag'}[J-a_1]}.
	\eqno\eq $$
The subscript $\lag'$ of $W$ means that we use
$\lag'$ here as the Lagrangian to calculate the generating functional
$W$. Since the state $\ket\psi$ satisfies the constraint
Eq.~\constraint, we have
$$ \at{{\widetilde\phi}_{\lag'}(x)}{J-a_1} = \at{\delta 
	W_{\lag'}[J]\over\delta J(x)}{J-a_1}
	=\bra\psi\phit(x)\ket\psi=0. \eqno\eq $$
Hence we have
$$ -\int V(\phi)=\at{\left[\int (J-a_1){\widetilde\phi}_{\lag'}
	-\int V_{\lag'}({\widetilde\phi}_{\lag'})\right]}
	{{\widetilde\phi}_{\lag'}
	=0}=-\int V_{\lag'}(0) \eqno\eq $$
which in turn leads to
$$  V(\phi) = V_{\lag'}(0). \eqn\ordord $$
(This is just another form of Jackiw's original result[\jackiw].)
Thus, the vacuum energy in the theory with Lagrangian
$\lag'$ is equal to the effective potential $V(\phi)$. Therefore, we
only need to sum over one-particle irreducible graphs with no external
fields. For any given number of loops, there are only a limited number
of such graphs.

The effective potential can be similarly defined for a composite
operator $O(x)$. (We only consider local operators $O(x)$. For
treatments of non-local operators, see [\cornwall].)
Adding an external source $\int d^dxJ(x)O(x)$ to the
system, we can define a functional of $J(x)$
$$ W[J] = \ln\int\left[{\cal D}\varphi\right] e^{-S+\int
	d^dxJ(x)O(x)} \eqno\eq $$
and its Legendre transform
$$\eqalign{ \sigma(x) =\ & {\delta W[J]\over\delta J(x)} \cr
	\Gamma[\sigma] =\ & \int d^dx J(x)\sigma(x) - W[J]. }
	\eqno\eq $$
When $\sigma(x)$ is a constant, the functional $\Gamma[\sigma]$ can be
written as
$$ \Gamma[\sigma] =\int d^dx U(\sigma). \eqno\eq $$
The function $U(\sigma)$ is called the effective potential for the
composite operator $O(x)$. As for the case of the ordinary
effective potential, we can show that the effective potential
$U(\sigma)$ is the minimal energy density of the system under
the constraint 
$$ \vev{O(x)} = \sigma. \eqn\compconst $$

We can introduce external sources for both $O(x)$ and
$\varphi(x)$ and define the effective potential $V(\sigma, \phi)$ by a
double Legendre transformation. Writing
$$ \lag(J) = \lag -JO, \eqn\lagj $$
we define
$$\eqalign{ \wtilde[J,h]=\ &\ln\int\left[{\cal D}\varphi\right]
	e^{-\int\lag+\int(JO+h\varphi)} \cr
	=\ & \ln\int\left[{\cal D}\varphi\right]
	e^{-\int\lag(J)+\int h\varphi}. } \eqno\eq $$
and its Legendre transform
$$ \gtilde[\sigma,\phi] =\int(J\sigma+h\phi)-\wtilde[J,h]\eqno\eq $$
where the new variables $\sigma(x)$ and $\phi(x)$ are defined by
$$\eqalign{ \sigma(x) =\ & {\delta\wtilde[J,h]\over\delta J(x)}, \cr
	\phi(x) =\ & {\delta\wtilde[J,h]\over\delta h(x)}. }
	\eqno\eq $$
In this paper we will transform with respect to $h$ first to get
$$ \Gamma'[J,\phi] = \int h\phi- \wtilde[J,h] 
	= \Gamma_{\lag(J)}[\phi]  \eqn\effj $$
Then we transform with respect to $J$, obtaining
$$\eqalign{ \gtilde[\sigma,\phi] =\ & \int J\sigma+\Gamma'[J,\phi], \cr
	\sigma(x) =\ & -{\delta\Gamma'[J,\phi]\over
	\delta J(x)}. } \eqn\secondapp $$
It is easy to see that these two approaches are equivalent to each
other.

IF $\sigma(x)$ and $\phi(x)$ are constant, $J(x)$ and $h(x)$ must be
also, and we can write
$$\eqalign{ \gtilde[\sigma, \phi] =\ & \int d^dx V(\sigma, \phi), \cr
	\Gamma'[J,\phi] =\ & \int d^dx V_{\lag(J)}(\phi). } \eqno\eq $$
where function $V_{\lag(J)}(\phi)$ is the ordinary effective
potential with $\lag(J)$ as the Lagrangian. From Eq.~\secondapp, we
find
$$ V(\sigma,\phi)=J\sigma+V_{\lag(J)}(\phi) \eqno\eq $$
If we treat $J$ as independent of $\sigma$ and $\phi$, we can consider
the right side of the above equation as a new function
$Y(\sigma,J,\phi)$ of three variables
$$ Y(\sigma,J,\phi)=J\sigma+V_{\lag(J)}(\phi). \eqn\defy $$ 
By the properties of Legendre transformation,
$V(\sigma,\phi)$ only depends on two independent variables $\sigma$
and $\phi$. Thus to get $V(\sigma,\phi)$ from $Y(\sigma,J,\phi)$, we
must have
$$ {\partial Y(\sigma,J,\phi)\over\partial J}=0. \eqn\minione $$
If we set the external source $h(x)$ for $\varphi(x)$ to be zero,
the function $V(\sigma, \phi)$ reduces to the effective potential
$U(\sigma)$ for the operator $O$. This condition is equivalent to
$$ {\partial V(\sigma, \phi)\over\partial\phi} = 0. \eqno\eq $$
The function $V(\sigma, \phi)$ is the minimal energy density among all
states that satisfy the constraints $\vev{O(x)}=\sigma$ and
$\vev{\varphi(x)}=\phi$, while the function $U(\sigma)$ is the minimal
energy density among states that satisfy only the single constraint
$\vev{O(x)}=\sigma$. Thus $U(\sigma)$ is the minimum of
$V(\sigma,\phi)$ for all values of $\phi$. Expressed in terms of the
function $Y(\sigma,J,\phi)$, the above condition becomes
$$ {\partial Y(\sigma,J,\phi)\over\partial\phi}=0. \eqn\minitwo $$
After we have solved Eqs.~\minione\ and \minitwo\ for $J(\sigma)$ and
$\phi(\sigma)$ from, the effective potential $U(\sigma)$ for
the composite operator is just
$$ U(\sigma)=Y\big(\sigma, J(\sigma), \phi(\sigma)\big). \eqn\ytou $$
In summary, we can define a function $Y(\sigma,J,\phi)$ by
Eq.~\defy. Using minimization conditions Eqs.~\minione\ and \minitwo,
we can determine the functions $J(\sigma)$ and
$\phi(\sigma)$. Substituting these into $Y(\sigma,J,\phi)$, we can
find out the effective potential $U(\sigma)$ as in Eq.~\ytou.

\chapter{Three-dimensional Examples}

In this section, we will use some examples in three dimensions to
illustrate our method. Let us consider a theory with Lagrangian
density
$$\eqalign{ \lag =\ & {1\over2}(\partial\varphi_1)^2+{1\over2}
	(\partial\varphi_2)^2+{1\over2}m^2(\varphi_1^2+\varphi_2^2)
	+{\lambda\over4}(\varphi_1^2+\varphi_2^2)^2
	+{\kappa\over6}(\varphi_1^2+\varphi_2^2)^3 \cr
	& + {\rm counterterms}. } \eqno\eq $$
In three dimensions, $\lambda$ has mass dimension one and $\kappa$ is
dimensionless. In the \msbar\ scheme, which we use, only the mass
counterterm, which is linear in $m^2$, depends on the mass
parameter. Thus the renormalized $\varphi^2$ operator is
$$ [\varphi^2]=2{\partial\lag\over\partial m^2}
	=\varphi_1^2+\varphi_2^2+{\rm counterterms}. \eqno\eq $$
We will use an external source $\half J[\varphi^2]$ that is
proportional to the mass term. Thus, $\lag(J)$ in Eq.~\lagj\ is the
original Lagrangian with $m^2$ replaced by $m^2-J$ and
$V_{\lag(J)}(\phi)$ is the ordinary effective potential with $m^2$
replaced by $m^2-J$. (In the following, we will use $\half\sigma$
instead of $\sigma$ in the formalism so that $\sigma$ corresponds to
the expectation value of $[\varphi^2]$.)

First we will find the ordinary effective potential
$V(\phi)$. Following the method that we discussed earlier, we shift
the fields by a constant amount
$$\eqalign{\varphi_1(x) =\ & \phit_1(x)+\phi, \cr
	\varphi_2(x) =\ & \phit_2(x). } \eqno\eq $$
The shifted Lagrangian without the terms linear in $\phit$ is
$$\eqalign{ \lag' =\ & {1\over2}(\partial\phit_1)^2
	+{1\over2}(\partial\phit_2)^2+{1\over2}(m^2+3\lambda\phi^2
	+5\kappa\phi^4)\phit_1^2+{1\over2}(m^2+\lambda\phi^2
	+\kappa\phi^4)\phit_2^2 \cr
	& +{1\over2}m^2\phi^2+{1\over4}\lambda\phi^4+{1\over6}
	\kappa\phi^6+\hbox{interaction terms}
	+\hbox{counterterms}. } \eqno\eq $$
To one-loop order, the effective potential is finite and equal to
$$\eqalign{ V(\phi)=&\ {1\over2}m^2\phi^2+{1\over4}\lambda\phi^4
	+{1\over6}\kappa\phi^6 \cr
	&\ -{1\over12\pi}\left[(m^2+3\lambda\phi^2
	+5\kappa\phi^4)^{3/2} + (m^2+\lambda\phi^2
	+\kappa\phi^4)^{3/2}\right]. } \eqn\vtoone $$
Replacing $m^2$ by $m^2-J$ in $V(\phi)$ gives $V_{\lag(J)}(\phi)$. The
function $Y(\sigma,J,\phi)$ is related to this by Eq.~\defy. To
one-loop order, we have 
$$\eqalign{ Y(\sigma,J,\phi) =\ & {1\over2}J\sigma+{1\over2}(m^2-J)
	\phi^2+{\lambda\over4}\phi^4+{\kappa\over6}\phi^6 \cr
	& -{1\over12\pi}\left[(m^2-J+3\lambda\phi^2+5\kappa\phi^4)^{3/2}
	+ (m^2-J+\lambda\phi^2+\kappa\phi^4)^{3/2}\right]. } \eqn\vone $$
Applying Eqs.~\minione\ and \minitwo\ to this function, we get
$$\eqalign{ \phi(m^2-J&+\lambda\phi^2+\kappa\phi^4)
	-{\phi\over4\pi}\left[(3\lambda+10\kappa\phi^2)
	(m^2-J+3\lambda\phi^2+5\kappa\phi^4)^{1/2}\right. \cr
	&\left. +(\lambda+2\kappa\phi^2)
	(m^2-J+\lambda\phi^2+\kappa\phi^4)^{1/2}\right] = 0
	} \eqn\threeone  $$
and
$$ \sigma-\phi^2+{1\over4\pi}\left[(m^2-J+3\lambda\phi^2
	+5\kappa\phi^4)^{1/2}+(m^2-J+\lambda\phi^2
	+\kappa\phi^4)^{1/2}\right] = 0. \eqn\threetwo $$

One solution to these equations is
$$\eqalign{ \phi =&\ 0 \cr (m^2-J)^{1/2}=&\ -2\pi\sigma.} \eqn\jfunc $$
Substituting this into $Y(\sigma, J, \phi)$ gives
$$ U(\sigma)=Y\big(\sigma,J(\sigma),\phi(\sigma)\big)
	={1\over2}m^2\sigma-{2\over3}\pi^2\sigma^3. \eqn\threesym $$
As we can see from Eq.~\jfunc, this is valid for $\sigma<0$. ($\sigma$
can be negative because we have subtracted a divergent number from it
to make it finite.) Since
$\phi=0$, the state corresponding to this solution is in the symmetric
phase with $\vev{\varphi(x)}=0$. In the above equation, the first term
is the classical value, and the second term comes from zero-point
energy of quantum oscillators around the origin.

A second, non-trivial, solution with $\phi\neq 0$ can be obtained by
solving the equations order by order. At tree-level, Eqs.~\threeone\
and \threetwo\ give
$$\eqalign{ \phi^2 =\ & \sigma, \cr
	J =\ & m^2+\lambda\sigma+\kappa\sigma^2.} \eqno\eq $$
Substituting these relations back into Eqs.~\threeone\ and \threetwo,
and keeping terms to one-loop order, we get
$$\eqalign{ \phi^2 =\ & \sigma+{1\over4\pi}(2\lambda\sigma
	+4\kappa\sigma^2)^{1/2}, \cr
	J =\ & m^2+\lambda\sigma+\kappa\sigma^2-{1\over2\pi}
	(\lambda+4\kappa\sigma)(2\lambda\sigma
	+4\kappa\sigma^2)^{1/2}. } \eqn\jphi $$
The effective potential in this case is
$$ U(\sigma)={1\over2}m^2\sigma+{1\over4}\lambda\sigma^2
	+{1\over6}\kappa\sigma^3-{1\over12\pi}(2\lambda\sigma
	+4\kappa\sigma^2)^{3/2}. \eqn\threeasym $$
This is valid for $\sigma>0$. Since $\phi\neq0$, the state
corresponding to this solution is in an asymmetric phase.

As $\sigma$ approaches zero from below and from above, Eq.~\threesym\
and \threeasym\ give
$$ \lim_{\sigma\rightarrow 0_-}U(\sigma) =
	\lim_{\sigma\rightarrow 0_+}U(\sigma) = 0 \quad\quad\quad
	\lim_{\sigma\rightarrow 0_-}{dU(\sigma)\over d\sigma}
	= \lim_{\sigma\rightarrow 0_+}{dU(\sigma)\over d\sigma}
	= {m^2\over2}. \eqno\eq $$
Therefore the effective potential and its first derivative are
continuous at where the symmetric and asymmetric solutions connect.

We have plotted the effective potential $U(\sigma)$ versus $\sigma$
and the ordinary effective potential $V(\phi)$ versus $\phi^2$ for the
case where $m^2<0$, $\lambda>0$ and $\kappa=0$ in Figure~1. The reason
that they look similar is that both are dominated by the
tree-level contributions, which are the same for both cases. There are
some important differences. Their one-loop order corrections are
different. More importantly, $U(\sigma)$ is real everywhere, while
$V(\sigma)$ has an imaginary part for small $\phi^2$ if
$m^2<0$. While $V(\phi^2)$ is only defined for $\phi^2>0$, the
effective potential $U(\sigma)$ is defined for all values of
$\sigma$. Moreover, the ordinary effective potential $V(\phi)$ has a
potential barrier for small $\phi$, while the effective potential
$U(\sigma)$ is a globally convex function without any potential
barrier in the case of $\varphi^4$ theory.

When $m^2>0$, $\kappa>0$ and $\lambda$ is an appropriately chosen
negative quantity, the ordinary effective potential $V(\sigma)$ has a
local minimum at $\phi=0$ and other local minima away from the
origin. In this case, the effective potential $U(\sigma)$ has two
minima, one corresponding to $\phi=0$, and the other to $\phi\neq0$,
as shown in Figure~2. The effective potential $U(\sigma)$ now has a
potential barrier between these two minima and it becomes complex in
this region. It is compared to the ordinary effective potential in
Figure~3.

Each point on $U(\sigma)$ and $V(\phi)$ represents a state. A point on
$U(\sigma)$ represents a state that has the minimal energy density among
all states that satisfy the constraint $\vev{[\varphi^2]}=\sigma$, and
a point on $V(\phi)$ represents a state that has the minimal energy
density among all states that satisfy the constraint
$\vev\varphi=\phi$. One may ask, for appropriately selected values of
$\sigma$ and $\phi$, whether $U(\sigma)$ and $V(\phi)$ represent the
same state. We shall now examine if there is a correspondence between
them. For $V(\phi)$, only the point $\phi=0$ corresponds to a state in
the symmetric phase. For $U(\sigma)$, all points that satisfy
$\sigma\leq\sigma_0$ for some value of $\sigma_0$ are in the symmetric
phase. Therefore, for $\sigma\leq\sigma_0$, except for the one point
of $U(\sigma)$ that corresponds to the point $\phi=0$ of $V(\phi)$,
no points of $U(\sigma)$ map to $V(\phi)$. Now
let us consider the asymmetric phase. For any value of $\sigma$,
$U(\sigma)$ is the minimal energy density among all states
that satisfy the constraint $\vev{[\varphi^2]}=\sigma$. Similarly, for
any value of $\phi$, $V(\phi)$ is the minimal energy
density among all states that satisfy the constraint
$\vev\varphi=\phi$. If the state represented by
$\vev{[\varphi^2]}=\sigma$ with minimal energy density $U(\sigma)$ has
expectation value $\vev\varphi=\phi$, we must have $V(\phi)\leq
U(\sigma)$, since $V(\phi)$ is the minimal energy density among all
states that satisfy the constraint $\vev\varphi=\phi$. For the state
represented by $\vev\varphi=\phi$ with minimal energy $V(\phi)$, its
expectation value $\vev{[\varphi^2]}=\sigma'$ is different from
$\sigma$ in general, unless the state represented by $\vev\varphi=\phi$
with energy density $V(\phi)$ is the same as the state represented by
$\vev{[\varphi^2]}=\sigma$ with energy density $U(\sigma)$. If these
two states are the same, there is a mapping between $U(\sigma)$ and
$V(\phi)$, and we have $\sigma=\sigma'$ and $U(\sigma)=V(\phi)$. If
they are not the same, then we must have $U(\sigma')<V(\phi)$ since
$U(\sigma')$ is the minimal energy density among all states which
satisfy the constraint $\vev{[\varphi^2]}=\sigma'$. Thus we conclude
that there is a mapping between $U(\sigma)$ and $V(\phi)$ if and only
if $\sigma'=\sigma$.

First, let us find $\vev\varphi=\phi$
for the state represented by $\vev{[\varphi^2]}=\sigma$ with minimal
energy density $U(\sigma)$. The argument $\phi$ in the function
$Y(\sigma, J, \phi)$ is the expectation value that we are
looking for. To one-loop order, the value of $\phi$ is given by
Eq.~\jphi.

We want to find the expectation value $\vev{[\varphi^2]}
=\sigma'$ for the state represented by $\vev\varphi=\phi$ with minimal
energy density $V(\phi)$. As we showed earlier, the renormalized
composite operator $[\varphi^2]$ is related to the Lagrangian by
$$ [\varphi^2]=2{\partial\lag\over m^2}. \eqno\eq $$
The state in question is a vacuum state for Lagrangian
$$ \lag'=\lag-J\varphi \eqno\eq $$
where $J$ is a parameter to be determined and does not dependent on
the space-time variable $x$. Hence
$$ \vev{[\varphi^2]} = { \int[{\cal D}\varphi] 2{\partial\lag\over
        \partial m^2}e^{-\int\lag+\int J\varphi} \over
        \int[{\cal D}\varphi]e^{-\int\lag+\int J\varphi} }.
        \eqno\eq  $$
Integrating this gives
$$\eqalign{ \int d^nx\vev{[\varphi^2]}=\ &
	{ -2{\partial\over\partial m^2}
        \int[{\cal D}\varphi]e^{-\int\lag+\int J\varphi}\over
        \int[{\cal D}\varphi]e^{-\int\lag+\int J\varphi} }
        = -2{\partial W[J]\over\partial m^2} \cr
	=\ & 2{\partial\Gamma[\phi]\over\partial m^2}
	=2\int d^nx {\partial V(\phi)\over\partial m^2}. } \eqno\eq $$
We conclude that
$$ \sigma'=\vev{[\varphi^2]}=2{\partial V(\phi)\over\partial
        m^2}. \eqn\sigmavalue $$

To one-loop order, the ordinary effective potential is given by
Eq.~\vtoone. Using Eq.~\sigmavalue\ and neglecting terms of
two-loop order or higher, we find that
$$ \sigma'=\phi^2-{1\over4\pi}\left[(m^2+3\lambda\phi^2
        +5\kappa\phi^4)^{1/2}+(m^2+\lambda\phi^2+\kappa\phi^4)
        ^{1/2} \right]. \eqn\ftos $$
Using Eqs.~\jphi\ and \ftos, we find that, to one-loop order, $\sigma$
and $\sigma'$ are related by
$$\eqalign{ \sigma' =\ & \sigma+{1\over4\pi}(2\lambda\sigma
        +4\kappa\sigma^2)^{1/2} \cr 
        &-{1\over4\pi}\left[(m^2+3\lambda\sigma+5\kappa\sigma^2)^{1/2}
        +(m^2+\lambda\sigma+\kappa\sigma^2)^{1/2}\right]. } \eqno\eq $$
At the local minima and local maxima of $U(\sigma)$, the equation
$$ m^2+\lambda\sigma+\kappa\sigma^2=0 \eqno\eq $$
holds at tree-level and $\sigma'=\sigma$.
We see that to one-loop order $\sigma'$ differs from $\sigma$ 
unless $m^2+\lambda\sigma+\kappa\sigma^2=0$ at tree-level. 
Therefore, we find that except at local minima and local
maxima, there is no mapping between $U(\sigma)$ and $V(\phi)$ in the
asymmetric phase.

\chapter{Ungauged $O(N)$ Model in Four Dimensions}

In three or fewer dimensions, the vacuum energy counterterm is either
linear in $m^2$ or independent of $m^2$. In four dimensions, however,
it is quadratic in $m^2$. Adding a source term
$\half J[\varphi^2]$ causes a divergence in the effective
potential because the vacuum energy divergence is not cancelled. In
this section, we will use an ungauged $O(N)$ model to study this
problem and find a solution.

The Lagrangian of our $O(N)$ model is
$$ \lag={1\over2}\partial_\mu\varphi_a\partial_\mu\varphi_a
	+{1\over2}m^2\varphi_a\varphi_a+{\lambda\over8N}
	(\varphi_a\varphi_a)^2+{\rm counterterms}. \eqn\lagon $$
Although we will not use the large-$N$ limit, we can use powers of $N$ to
organize our results.

Of all the counterterms, only the mass and vacuum energy counterterms
depend on $m^2$, and they depend on it in a simple way (so-called
soft-parameterization[\collins]). To all orders, we can write the mass
terms as
$$ {1\over2}m^2\varphi_a\varphi_a\left(1+\sum_{i=1}^\infty
	b_i\lambda^i\right) \eqno\eq $$
where $b_i$'s are simple poles in $4-d$. Similarly, the vacuum energy
counterterms can be written as
$$ {m^4\over4}\sum_{i=1}^\infty c_i\lambda^{i-1} \eqno\eq $$
where $c_i$'s are also simple poles in $4-d$. All other terms are
independent of $m^2$ in our \msbar\ scheme.

By differentiating the Lagrangian with respect to the renormalized
mass parameter $m^2$, we can get the renormalized composite operator
$\phir$:
$$ \phir=2{\partial\lag\over m^2}=\varphi_a\varphi_a\left(1+\sum
	b_i\lambda^i\right)+m^2\sum c_i\lambda^{i-1}. \eqno\eq  $$
This operator is finite in the sense that all matrix elements with one
insertion of this operator are finite. However, matrix elements with
more than one insertion of this operator are in general
divergent. This can be seen from the fact that the generating 
functionals we get by adding external sources coupled to this operator
are divergent.

Consider the Lagrangian obtained by replacing $m^2$ in Eq.~\lagon\
with $m^2-J$. Only the mass and vacuum energy terms are 
affected by this replacement. The mass term becomes
$$ {1\over2}(m^2-J)\varphi_a\varphi_a\left(1+\sum b_i\lambda^i
	\right), \eqno\eq $$
and the vacuum term becomes
$$ {(m^2-J)^2\over4}\sum c_i\lambda^{i-1}. \eqno\eq $$
The Lagrangian differs from the original one by a term linear in $J$
$$ -{1\over2}J\left[\varphi_a\varphi_a\left(1+\sum b_i\lambda^i
	\right)+m^2\sum c_i\lambda^{i-1}\right]
	=-{1\over2}J\phir \eqn\rsource $$
and a term quadratic in $J$
$$ {J^2\over4}\sum c_i\lambda^{i-1}. \eqn\quadterm $$
Therefore, if we added these two terms as sources to the original
Lagrangian, we would get a finite theory with mass parameter
$m^2-J$. However, we are not allowed to add source terms quadratic in
$J$. If we only add the source term of Eq.~\rsource, the generating
function $W[J]$ will be divergent because of the lack of the divergent
$J^2$ terms in Eq.~\quadterm. Consequently, the effective potential
for the operator $\phir$ will be divergent.

In order for the generating functional $W[J]$ to be finite, the matrix
elements with any number of insertions of the source term must be
finite. It is not possible to find such an composite operator for
all states. However, the effective potential is the minimal energy
density under the given constraint. We only need to find a composite
operator such that for the state with the minimal energy density, the
matrix elements with any number of insertions of this operator are
finite. If we add extra terms to the composite operator $\phir$, it
is possible to cancel the $J^2$ divergence at the minimizing point
only. So we define a new composite operator
$$ \phis=\phir+\sum f_i(\varphi_a\varphi_a). \eqno\eq $$
With this new operator, the function $Y(\sigma, J, \phi)$ is still
divergent. The effective potential $U(\sigma)$ obtained by applying
the minimization conditions Eqs.~\minione\ and \minitwo\ to $Y(\sigma,
J, \phi)$ will be finite. We will adjust the coefficient functions
$f_i$'s order by order in perturbation theory so that the
divergences in the effective potential for the composite operator
$\phis$ are cancelled.

Adding a source
$$ -{1\over2}J\phis \eqn\source $$
to the original $O(N)$ Lagrangian $\lag(\varphi; m^2)$ with mass
parameter $m^2$, gives a new Lagrangian
$$ \lag(J) = \lag(\varphi; m^2-J)-{1\over2}J\sum\left[
	f_i(\varphi_a\varphi_a)+{1\over2}Jc_i\lambda^{i-1}\right].
	\eqn\modlag $$
The ordinary effective potential then becomes
$$ V_{\lag(J)}(\phi)=V(\phi;m^2-J)+\vex(J,\phi) \eqno\eq  $$
where $V(\phi; m^2-J)$ is the ordinary effective potential but
with $m^2-J$ as its mass parameter and $\vex(J,\phi)$ arises from
the last term in Eq.~\modlag, either directly or through insertion
into a larger graph. From $V_{\lag(J)}(\phi)$, we obtain
$$ Y(\sigma, J,\phi)={1\over2}J\sigma+V_{\lag(J)}(\phi). \eqno\eq $$
We will adjust the $f_i$'s so that after minimization with respect to
$J$ and $\phi$, the function $Y(\sigma,J,\phi)$ yields a finite
effective potential $U(\sigma)$.

We must first obtain the ordinary effective potential and some of the
counterterms. Shifting the fields by
$\varphi_a(x)=\phit_a(x)+\phi\delta_{aN}$, we find that, up to
one-loop order in the \msbar\ scheme, the effective potential is
$$\eqalign{ V(\phi)=\ &{1\over2}m^2\phi^2+{\lambda\over8N}\phi^4
	+{\left(m^2+{3\lambda\over2N}\phi^2\right)
	^2\over64\pi^2}\left(\ln{m^2+{3\lambda\over2N}\phi^2\over
	\mubar}-{3\over2}\right) \cr &+(N-1){\left(m^2+{\lambda
	\over2N}\phi^2\right)^2\over64\pi^2}\left(\ln{m^2+{\lambda
	\over2N}\phi^2\over\mubar}-{3\over2}\right). } \eqno\eq $$
The one-loop order vacuum energy counterterm is
$$ c_1={N\over 16\pi^2\epsilon}. \eqn\cone $$
The two-loop order contributions can be similarly calculated. The
two-loop order vacuum energy counterterm is
$$ c_2={(N+2)\over2(4\pi)^4\epsilon^2}. \eqn\ctwo $$

Now we are ready to study the effective potential for $\phis$. To
one-loop order, we have
$$\eqalign{ \ & Y(\sigma,J,\phi^2)  \cr =\ &{1\over2}J\sigma
	+{1\over2}(m^2-J)\phi^2+{\lambda\over8N}\phi^4
	+{\left(m^2-J+{3\lambda\over2N}\phi^2\right)^2\over64\pi^2}
	\left(\ln{m^2-J+{3\lambda\over2N}\phi^2\over\mubar}-{3\over2}\right)
	\cr&+(N-1){\left(m^2-J+{\lambda\over2N}\phi^2\right)^2\over64\pi^2} 
	\left(\ln{m^2-J+{\lambda\over2N}\phi^2\over\mubar}-{3\over2}\right)
	-{1\over2}J\left[f_1(\phi^2)+{1\over2}Jc_1\right]. } \eqno\eq $$
(For convenience, we will use $\phi^2$ instead of $\phi$ as one of
$Y$'s argument.) At tree-level, the minimization conditions
Eqs.~\minione\ and \minitwo\ give $J_0 = m^2+{\lambda\over2N}\sigma$
and $\sigma(\phi^2)_0=\sigma$. Substituting this back into $Y(\sigma,
J, \phi^2)$, we find that to one-loop order the effective potential is
$$\eqalign{ U(\sigma)=\ &Y_0\big(\sigma,J_0,(\phi^2)_0\big)
	+\attree{\partial Y_0\over\partial J}J_1
	+\attree{\partial Y_0\over\partial(\phi^2)}(\phi^2)_1
	+Y_1\big(\sigma, J_0,(\phi^2)_0\big) \cr
	=\ &{1\over2}m^2\sigma+{\lambda\over8N}\sigma^2
	+{1\over64\pi^2}\left({\lambda\sigma\over N}\right)^2
	\left(\ln{\lambda\sigma\over N\mubar}-{3\over2}\right)
	-{1\over2}J_0\left[f_1(\sigma)+{1\over2}J_0c_1\right] \cr
	=\ &{1\over2}m^2\sigma+{\lambda\over8N}\sigma^2
	+{1\over64\pi^2}\left({\lambda\sigma\over N}\right)^2
	\left(\ln{\lambda\sigma\over N\mubar}-{3\over2}\right). }
	\eqn\uonone $$
The second and third terms on the first line vanish because of the
minimization conditions Eqs.~\minione\ and \minitwo. 
To make $U(\sigma)$ finite to this order, we must have
$$ f_1(\sigma)=-{1\over2}J_0c_1 = -{N\over32\pi^2\epsilon}
	\left(m^2+{\lambda\over2N}\sigma\right). \eqno\eq $$
(There is no finite term in above equation because of the \msbar\
scheme we use.) Using this $f_1$, we apply the minimization conditions
Eqs.~\minione\ and \minitwo\ and find that to one-loop order
$$\eqalign{ J=\ & J_0+J_1=m^2+{\lambda\sigma\over2N}
	+{\lambda^2\sigma\over16\pi^2N^2}\left(\ln
	{\lambda\sigma\over N\mubar}-1\right), \cr
	(\phi^2)=\ &(\phi^2)_0+(\phi^2)_1=
	\sigma-{\lambda\sigma\over16\pi^2N}\left(\ln
	{\lambda\sigma\over N\mubar}-1\right)-{N\over32\pi^2\epsilon}
	\left(m^2+{\lambda\over2N}\sigma\right). } \eqno\eq $$
Notice that $U$ and $J$ are finite functions of $\sigma$, while
$(\phi^2)$ is a divergent function of $\sigma$. 

Expanding $Y$ to two-loop order, we find that the two-loop order
contribution to the $U(\sigma)$ is
$$\eqalign{ U_2(\sigma)=\ &
	{1\over2}J_1(\phi^2)_1-{\lambda\over8N}
	(\phi^2)_1^2+V_2\big((\phi^2)_0; m^2-J_0\big) \cr
	&-{1\over2}\left(m^2+{\lambda\over2N}\sigma\right)\left[
	f_2(\sigma)+{1\over2}\lambda c_2\left(m^2+{\lambda\over2N}
	\sigma\right)\right]+\uex'(\sigma). } \eqn\twoeff $$
where $c_2$ is the two-loop order vacuum energy counterterm
coefficient given in Eq.~\ctwo\ and the $\uex'(\sigma)$ term comes from
insertions of 
$-{1\over2}Jf_1(\varphi_a\varphi_a)$ in one-loop graphs:
$$ \uex'(\sigma) = {\lambda^2\sigma\over8N(4\pi)^4}
	\left(m^2+{\lambda\over2N}\sigma\right)
	\left[-{1\over\epsilon^2}+{1\over\epsilon}
	\left(\ln{\lambda\sigma\over N\mubar}-1
	\right)\right]+\hbox{finite terms}. \eqno\eq $$
Hence
$$\eqalign{ U_2(\sigma)= -{1\over2}\left(m^2+{\lambda\over2N}\sigma\right)
	&\left\{f_2(\sigma)+{\lambda^2\sigma\over2N(4\pi)^4\epsilon}
	\left(\ln{\lambda\sigma\over N\mubar}-1\right) \right. \cr
	&\left. +{\lambda\over4(4\pi)^4\epsilon^2}\left[
	{\lambda\sigma\over N}+\left({5\over4}N+2\right)\left(
	m^2+{\lambda\sigma\over2N}\right)\right]\right\}
	+\hbox{finite terms} } \eqno\eq $$
To make this finite, we must have
$$f_2(\sigma)= -{\lambda\over4(4\pi)^4\epsilon^2}\left[
	{\lambda\sigma\over N}+\left({5\over4}N+2\right)
	\left(m^2+{\lambda\sigma\over2N}\right)\right]
	-{\lambda^2\sigma\over2N(4\pi)^4\epsilon}
	\left(\ln{\lambda\sigma\over N\mubar}-1\right) \eqno\eq $$
in our minimal subtraction scheme. Notice that the function
$f_2(\varphi_a\varphi_a)$ not only has terms proportional to $m^2$ and
$\varphi_a\varphi_a$, but also has terms logarithmic in
$\varphi_a\varphi_a$. Thus, beginning at two-loop order, the
counterterms in $\phis$ become non-polynomial.

Our method is only applicable for the asymmetric solution. In that
case, both $J$ and $\phi$ vary with $\sigma$, so adding
appropriate counterterms can cancel the divergence proportional to
$J^2$. For the symmetric solution, $\phi$ is a constant and so we
cannot cancel this divergence by adding
counterterms. We have been unable to find a way to define a finite
composite effective potential in the symmetric phase.

\chapter{Scalar QED in Four Dimensions}

In this section, we will use the method demonstrated above to study
the effective potential for $\phis$ for scalar QED in four dimensions.
As in the case of ungauged $O(N)$ model in four dimensions, we will
need to add extra counterterms to the conventionally defined operator
$\phir$ to obtain a finite composite effective potential. We will also examine
the gauge dependence of the composite effective potential.

The Lagrangian of scalar QED is
$$ \lag={1\over2}D_\mu\varphi_aD_\mu\varphi_a
	+{1\over4}F_{\mu\nu}F_{\mu\nu}+{1\over2}m^2\varphi^2
	+{\lambda\over4}\varphi^4
	\eqno\eq $$
where $\varphi^2=\varphi_1^2+\varphi_2^2$, $\varphi^4=(\varphi^2)^2$
and $\varphi_a(a=1,2)$ are real fields. We will use the \rxi\ with a
gauge fixing term
$$ {1\over2\xi}(\partial\cdot A+ev\cdot\varphi)^2 \eqno\eq $$
where $v_a$ is an external 2-vector. This gauge fixing term requires a
ghost compensating term
$$ \partial_\mu c^\ast\partial_\mu c+e^2(v\times\varphi)c^\ast c.
	\eqno\eq $$

The theory has two dimensionful parameters, $m^2$ and $v$. Of the
counterterms, only the mass and the vacuum energy
counterterms depend on $m^2$. We can write these as
$$ {1\over2}m^2\varphi_a\varphi_a\left(1+\sum_{i=1}^\infty
	b_i(\lambda,e^2)\right)+\hbox{$m^2$-independent terms},
	\eqno\eq $$
and
$$ {m^4\over4}\sum_{i=1}^\infty c_i(\lambda,e^2)
	+{1\over2}m^2v^2\sum_{i=1}^\infty d_i(\lambda, e^2)
	+\hbox{$m^2$-independent terms} \eqno\eq $$
where the $b_i$'s are polynomials of $\lambda$ and $e^2$ of order $i$, and
the $c_i$'s and $d_i$'s are polynomials of $\lambda$ and $e^2$ of order
$i-1$. The coefficients of these polynomials are simple poles in
$4-d$.

As in the case of the ungauged $O(N)$ model in four
dimension, the effective potential for the conventionally defined
composite operator $\phir$ is divergent. We need to use a composite
operator with additional counterterms,
$$ \phis=\phir+\sum f_i(\varphi_a\varphi_a). \eqno\eq $$
With this coupled to a source, we have
$$ \lag(J)=\lag(\varphi,A,c^\ast,c;m^2-J)-{1\over2}J\sum
	\left[f_i(\varphi_a\varphi_a)+{1\over2}Jc_i\right].
	\eqn\qedlag $$

To calculate the ordinary effective potential $V(\phi;m^2)$, we
shift the scalar fields by a constant amount, $\varphi_a = \phit_a
+\delta_{a1} \phi$.
To be consistent, the $v$-vector has to be chosen as $v_a=\delta_{a2}v$.
Up to one-loop order, the renormalized ordinary effective potential is
$$\eqalign{ V(\phi) =\ &{1\over2}m^2\phi^2+{\lambda\over4}\phi^4
	+{1\over4(4\pi)^2}\left[(m_1^2)^2\left(\ln{m_1^2\over\mubar}
	-{3\over2}\right)+3(e^2\phi^2)^2\left(\ln{e^2\phi^2\over\mubar}
	-{5\over6}\right) \right. \cr
	& \left. -2(e^2v\phi)^2\left(\ln{-e^2v\phi\over\mubar}
	-{3\over2}\right)+r_1^2\left(\ln{r_1\over\mubar}-{3\over2}
	\right)+r_2^2\left(\ln{r_1\over\mubar}-{3\over2}\right)
	\right] } \eqno\eq $$
where $m_1^2=m^2+3\lambda\phi^2$, $m_2^2=m^2+\lambda\phi^2$, and $-r_1$
and $-r_2$ are the roots of $(k^2/\xi+e^2\phi^2)(k^2+m_2^2+e^2v^2/\xi)
-e^2k^2(\phi+v/\xi)^2$.
The one- and two-loop order vacuum energy counterterms are
$$ {m^4\over4}c_1 = {m^4\over2(4\pi)^2\epsilon} \eqno\eq $$
and
$$ {m^4\over4}c_2+{m^2v^2\over2}d_2 = 
	\left({m\over4\pi}\right)^4\left[{2\lambda\over\epsilon^2}
	-{3e^2\over2\epsilon^2}+{2e^2\over\epsilon}\right]
	+{e^4m^2v^2\over(4\pi)^4}\left[{3\over4\epsilon^2}
	+{1\over4\epsilon}\right]. \eqn\twovac $$

Proceeding as in Sec.~{\Roman 4}, we find that to one-loop order the effective potential for the composite operator is
$$ U(\sigma)={1\over2}m^2\sigma+{\lambda\over4}\sigma^2
	+{1\over4(4\pi)^2}\left[4\lambda^2\sigma^2\left(
	\ln{2\lambda\sigma\over\mubar}-{3\over2}\right)+3e^4\sigma^2
	\left(\ln{e^2\sigma\over\mubar}-{5\over6}\right)\right],
	\eqno\eq $$
while the function $f_1$ is
$$ f_1(\varphi_a\varphi_a)=-{1\over(4\pi)^4\epsilon}
	\left(m^2+\lambda\varphi_a\varphi_a\right). \eqno\eq $$
The relationship between $J$ and $\sigma$ and that between $\phi$ and
$\sigma$ are
$$\eqalign{ J(\sigma)=\ & m^2+\lambda\sigma+{1\over(4\pi)^2}
	\left[4\lambda^2\sigma
	\left(\ln{2\lambda\sigma\over\mubar}-1\right)+3e^4\sigma
	\left(\ln{e^2\sigma\over\mubar}-{1\over3}\right)\right] \cr
	=\ & J_0+J_1, } \eqno\eq $$
and
$$\eqalign{ (\phi^2)=\ &\sigma-{1\over(4\pi)^2}\left[2\lambda\sigma
	\left(\ln{2\lambda\sigma\over\mubar}-1\right)+{m^2+\lambda
	\sigma\over\epsilon} \right. \cr & \left.
	-(\xi e^2\sigma+2e^2v\sqrt{\sigma})\left(\ln{-e^2v
	\sqrt{\sigma}\over\mubar}-1\right)\right] \cr
	=\ & (\phi^2)_0+(\phi^2)_1 }  \eqno\eq $$
The two-loop order
correction to the effective potential can be written as
$$ U_2(\sigma) = {1\over2}J_1(\phi^2)_1-{\lambda\over4}(\phi^2)_1^2
	+Y_2\big(\sigma,J_0,(\phi^2)_0\big), \eqno\eq $$
where
$$ Y_2(\sigma, J, \phi^2) = V_2(\phi^2;m^2-J)-{J\over2}\left[
	f_2(\phi^2)+{1\over2}Jc_2\right]+\vex(J,\phi^2)
	\eqno\eq $$
The term $\vex(J,\phi^2)$, from insertions of $f_1$ in one-loop order
in one-loop graphs, is
$$ {\lambda J\over2(4\pi)^2\epsilon}\left(
        \dk{1\over k^2+m_1^2}+\dk{k^2+\xi e^2\phi^2\over(k^2+r_1)
        (k^2+r_2)}\right), \eqn\extrat $$
while $C_2$ is given in Eq.~\twovac.
The total divergent part of $U_2(\sigma)$ is
$$\eqalign{ &{\lambda^2\sigma\over\epsilon}\left(-2\ln{2\lambda\sigma
        \over\mubar}+\ln{-e^2v\sqrt{\sigma}\over\mubar}\right)
        -{3e^4\sigma\over2\epsilon}\ln{e^2\sigma\over\mubar}
        -{9\lambda m^2\over4\epsilon^2}+e^2m^2\left({3\over\epsilon^2}
        -{2\over\epsilon}\right) \cr
        + & \lambda^2\sigma\left(-{13\over4\epsilon^2}
        +{2\over\epsilon}\right)+{e^4\sigma\over2\epsilon}
        +\lambda e^2\sigma\left({3\over\epsilon^2}-{2\over\epsilon}
        \right)+\xi\lambda e^2\sigma\left({1\over 2\epsilon^2}
        -{1\over2\epsilon}\right)+{\lambda d^2v\sqrt{\sigma}
        \over\epsilon^2} \cr
        - & {(4\pi)^4\over2}f_2(\sigma). } \eqno\eq $$
This divergent part is zero in \msbar\ scheme. This condition
determines the function $f_2$ uniquely. With this $f_2$, the two-loop
order correction to the effective potential, $U_2(\sigma)$, is finite:
$$ U_2(\sigma)=V_2\big((\phi^2)_0; m^2-J_0\big)+G_1+G_2 \eqno\eq $$
where $G_1$ is the finite part of ${1\over2}J_1(\phi^2)_1
-{\lambda\over4}(\phi^2)_1^2$ and $G_2$ is finite part of Eq.~\extrat.

Let us examine the gauge dependence of $U_2(\sigma)$. To calculate
${\partial V_2\over\partial\xi}$, we will utilize the Nielsen
identity[\nielsen]
$$ \xi{\partial V\over\partial\xi}=C(J, \phi^2, \xi){\partial V\over
        \partial\xi}. \eqno\eq $$
Since the leading order of the function $C(J,\phi^2,\xi)$ is one-loop
order, we have
$$\eqalign{ \xi{\partial V_0\over\partial\xi} =\ & 0, \cr
        \xi{\partial V_1\over\partial\xi} =\ & C_1{\partial V_0
        \over\partial \phi}, \cr
        \xi{\partial V_2\over\partial\xi} =\ & C_1{\partial V_1
        \over\partial \phi} + C_2{\partial V_0\over\partial\phi}. }
        \eqn\nielsenseries $$
Using the second equation in Eq.~\nielsenseries\ and take the limit of
$J=J_0$ and $\phi^2=\phi^2_0$, we get
$$ C_1(J_0, \phi^2_0, \xi) = - {\xi e^2\sqrt{\sigma}\over2(4\pi)^2}
        \left(\ln{-e^2v\sqrt{\sigma}\over\mubar}-1\right).
        \eqno\eq $$
We want to evaluate ${\partial V_2\over\partial\xi}$ at the point
$J=J_0$ and $\phi^2=(\phi^2)_0$. At this point we have
$$ \attree{\partial V_0\over\partial\phi} = 0 . \eqno\eq $$
Using the result for $C_1$, we have
$$\eqalign{ \attree{\partial V_2\over\partial\xi} = &\ 
	{1\over\xi}C_1(J_0,\phi^2_0, \xi)
        \attree{\partial V_1\over\partial\phi} \cr
        =&\ -{e^2\sigma\over2(4\pi)^4} 
        \left(\ln{-e^2v\sqrt{\sigma}\over\mubar}-1\right)
        \left[6\lambda^2\sigma\left(\ln{2\lambda\sigma\over\mubar}
        -1\right)+3e^4\sigma\left(\ln{e^2\sigma\over\mubar}
        -{1\over3}\right) \right. \cr
        &\ \left. -e^2\lambda(\xi\sigma+2v\sqrt{\sigma})
        \left(\ln{-e^2v\sqrt{\sigma}\over\mubar}-1\right)
        \right]. } \eqno\eq $$
The gauge dependence of $G_1$ is
$$\eqalign{ {\partial G_1\over\partial\xi} =\ & {1\over2}\left[J_1
        -\lambda(\phi^2)_1^{\rm fin}\right]{\partial
        (\phi^2)_1^{\rm fin}\over\partial\xi} \cr
        =\ & {e^2\sigma\over2(4\pi)^4}\left(\ln{-e^2v\sqrt{\sigma}
        \over\mubar}-1\right)\left[6\lambda^2\sigma\left(
        \ln{2\lambda\sigma\over\mubar}-1\right) \right. \cr
        & \left. +3e^4\sigma\left(\ln{e^2\sigma\over\mubar}
        -{1\over3}\right)-\lambda(\xi e^2\sigma+2e^2v\sqrt{\sigma})
        \left(\ln{-e^2v\sqrt{\sigma}\over\mubar}-1\right)\right]. }
        \eqno\eq $$
As we can see
$$ \attree{\partial V_2\over\partial\xi}+{\partial G_1\over
        \partial\xi} = 0. \eqn\gaugeind $$
This is a consequence of the Nielsen identity applied to the ordinary
effective potential. To see this, let us define
$$ \ytilde(\sigma, J, \phi^2) = {1\over2}J\sigma +V(\phi^2;m^2-J).
	\eqno\eq $$
We find that the solution $J_0+J_1$ and $\phi^2_0+(\phi^2)_1^{\rm
fin}$ minimizes $\ytilde(\sigma, J, \phi^2)$ to two-loop order. At
this point,
$$ {\partial\ytilde\over\partial\xi}={\partial\ytilde\over\partial J}
	{\partial J\over\partial\xi} + {\partial\ytilde\over
	\partial\phi^2}{\partial \phi^2\over\partial\xi}
	+{\partial\ytilde\over\partial\xi}
	= {\partial V\over\partial\xi}. \eqno\eq $$
By Nielsen identity, at this point we have
$$ {\partial V\over\partial\xi} = {C\over\xi}{\partial V
	\over\partial\phi} = 0. \eqno\eq $$
Therefore $\ytilde$ at $J=J_0+J_1$ and
$\phi^2=\phi^2_0+(\phi^2)_1^{\rm fin}$ are gauge independent to
two-loop order. Thus, the two-loop order contribution to $\ytilde$,
$$ V_2+G_1=V_2(\phi^2_0;m^2-J_0)+{1\over2}J_1(\phi^2)_1^{\rm fin}
	-{\lambda\over4}\left[(\phi^2)_1^{\rm fin}\right]^2,
	\eqno\eq $$
is gauge independent as shown in Eq.~\gaugeind.
So the gauge dependence of $U(\sigma)$ at two-loop order comes from
$G_2$, which is the finite part of insertions of $f_1$ in one-loop
graphs. The term ${\partial G_2\over\partial\xi}$ is the finite part of
$$ {\lambda(m^2+\lambda\sigma)\over2(4\pi)^2\epsilon}\dk 
        { e^2\sigma\over(k^2-e^2v\sqrt{\sigma})^2}. \eqno\eq $$
We find that the finite part is
$$\eqalign{ {\partial G_2\over\partial\xi} = {\lambda e^2\sigma
        (m^2+\lambda\sigma)\over2(4\pi)^4} & \left[{1\over2}
        \ln^2{-e^2v\sqrt{\sigma}\over\mu^2}+{1\over2}\ln^2 4\pi
        +\beta \right. \cr
        & \left. -\gamma\ln{-e^2v\sqrt{\sigma}\over 4\pi\mu^2}
        -\ln4\pi\ \ln{-e^2v\sqrt{\sigma}\over\mu^2}\right]. }
        \eqno\eq $$
For the composite effective potential, we have
the following result
$$ {\partial U_2(\sigma)\over\partial\xi} = 
        {\partial G_2\over\partial\xi} \neq 0. \eqno\eq $$
When the extra counterterms to the operator $\phis$ at one-loop order
are inserted to one-loop graphs, they cause gauge-dependent
contributions to the effective potential $U(\sigma)$ at two-loop level.

However, we can modify our scheme to make the effective potential
$U(\sigma)$ gauge independent. We will add and adjust finite terms to
the operator $\phis$ order by order in the perturbation expansion. In
the scheme where finite parts vanish, we have shown that the effective
potential $U(\sigma)$ is gauge independent at tree and one-loop
level. Suppose in zero-finite-part scheme, at all levels of $n-1$
loops and less, the effective potential is gauge independent, and at
$n$-loop order the effective potential becomes gauge dependent. We can
add a finite counterterm $F_n(\varphi_a\varphi_a)$ to the operator
$\phis$. In this new scheme, the effective potential becomes
$$ U(\sigma)+F_n(\sigma) \eqno\eq $$
at order $n$. We can choose the function $F_n$ to cancel any
gauge-dependent piece of the effective potential $U(\sigma)$ to make
the effective potential gauge independent. We can go on to carry out
this procedure at higher orders. In this new scheme, the effective
potential will be gauge independent. We must stress that since one can
always add finite terms to make any gauge-dependent quantities
gauge-independent, and there is no preferred prescription for choosing
the finite part, this new modified scheme is not very useful in
practice.

\chapter{Conclusion}

We have demonstrated our method of calculating the composite effective
potential in three and four dimensions. It is straightforward to
generalize our method to different number of dimensions. In one
dimension, the operator $\varphi^2$ is finite and we can use it
directly in our calculation of composite effective potential. In two
and three dimension, this operator becomes divergent and we need to
use the renormalized operator by subtracting a divergent quantity from
this operator. It is easy to see that the composite effective
potential is gauge independent in three or fewer dimensions because of
the Nielsen identity. However, in four dimensions, there is no finite
effective potential for the conventionally defined composite operator
$\phir$ because graphs with two insertions of this operator remain
divergent. Nevertheless, we find that a finite effective potential
exists for a modified composite operator in four
dimensions. The modified composite operator is the sum of the
conventionally defined renormalized operator and some new
counterterms. By adjusting the counterterms order by order in a
perturbative scheme, we can make the composite effective potential
finite. However, the counterterms in the new operator $\phis$ are no
longer purely polynomial in the elementary fields. In a scheme where
all counterterms are pure poles in $4-d$, this finite effective
potential is gauge dependent because of the extra counterterms in
$\phis$. We have shown the gauge dependence explicitly at two-loop
order.

I would like to thank Erick Weinberg for suggesting this subject and
for numerous discussions.

\refout

\vfill\eject
\smallskip
\epsfxsize=\hsize
\epsfbox[12 12 576 530]{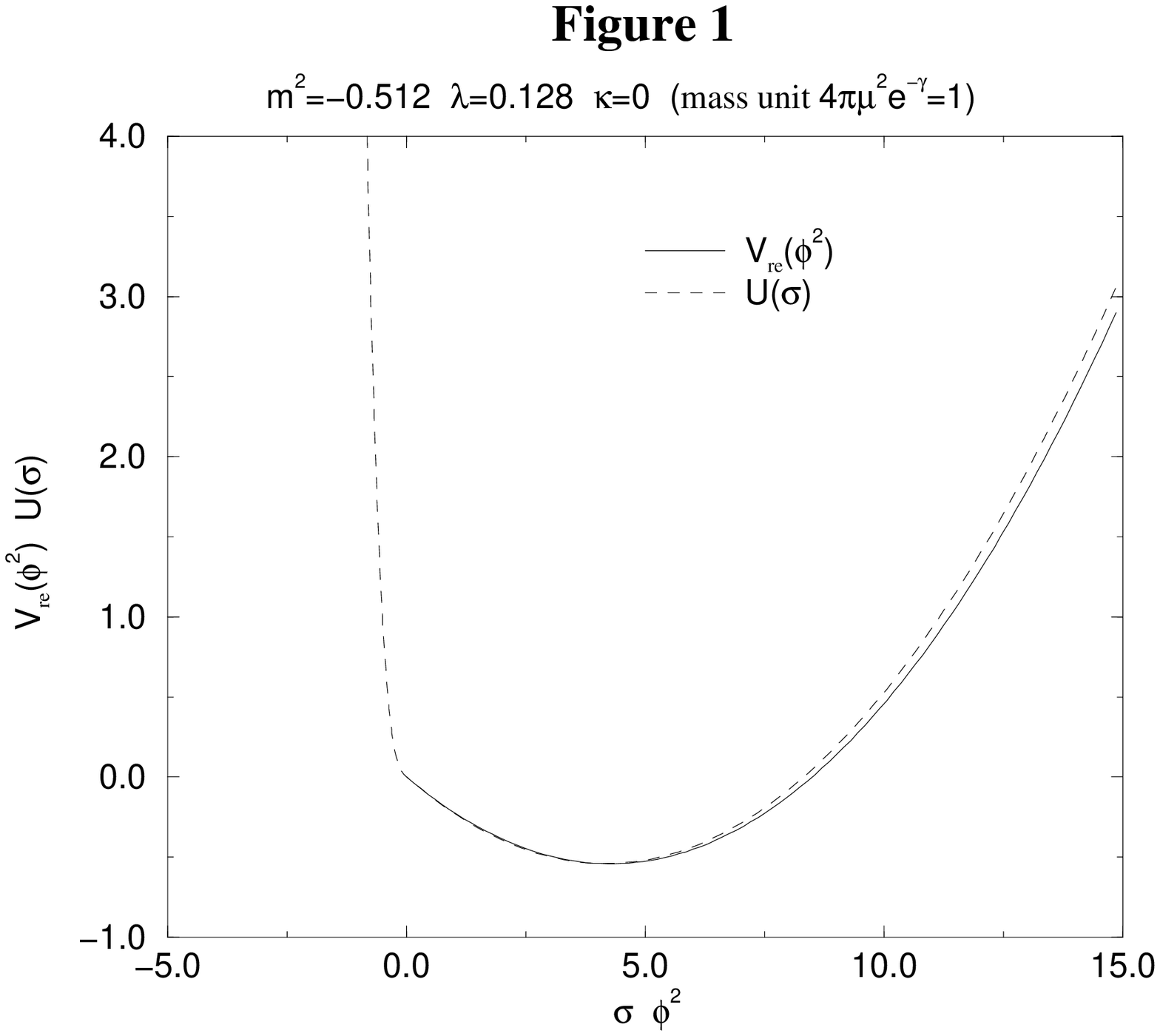}

\vfill\eject
\smallskip
\epsfxsize=\hsize
\epsfbox[12 12 576 530]{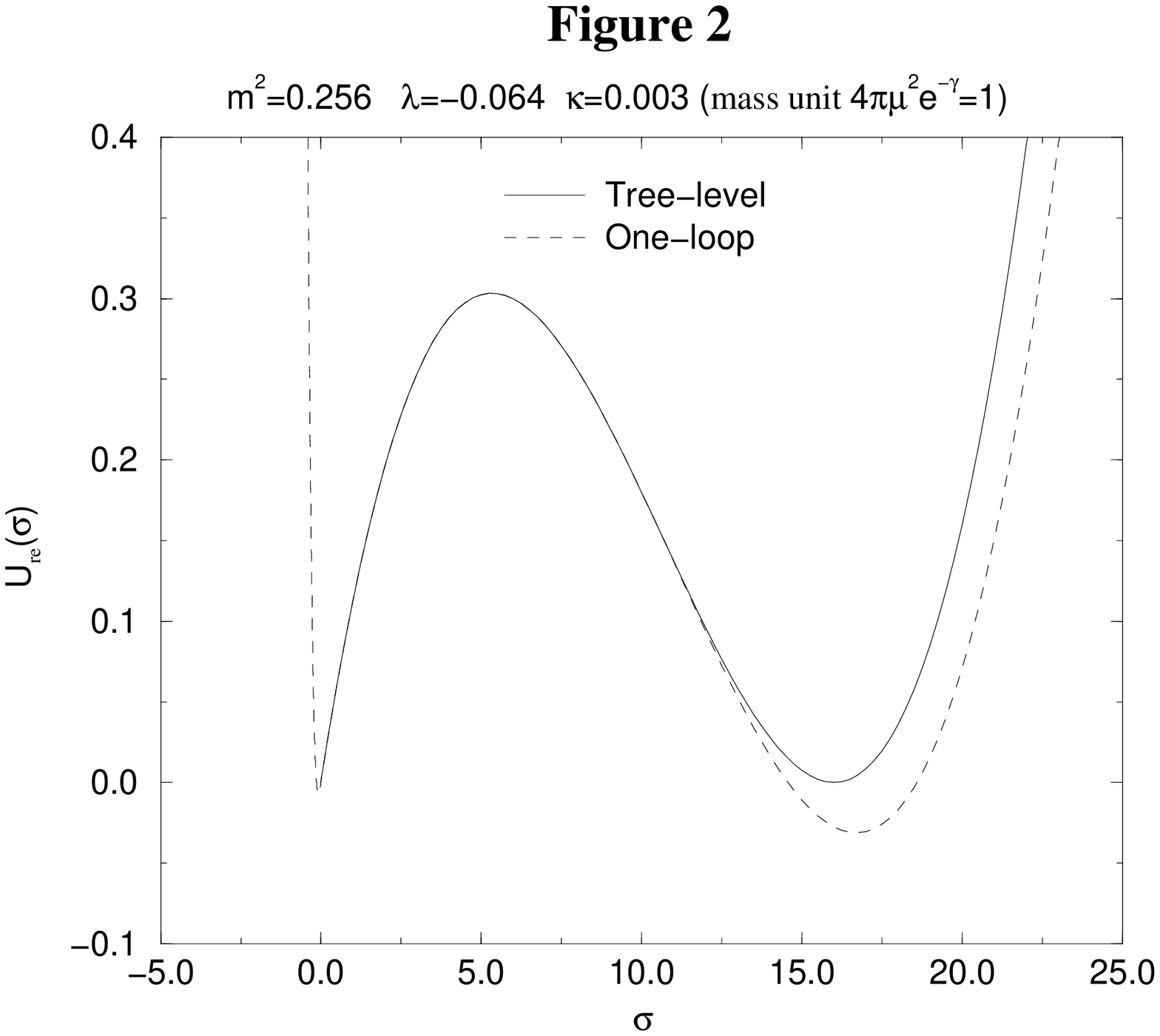}

\vfill\eject
\smallskip
\epsfxsize=\hsize
\epsfbox[12 12 576 530]{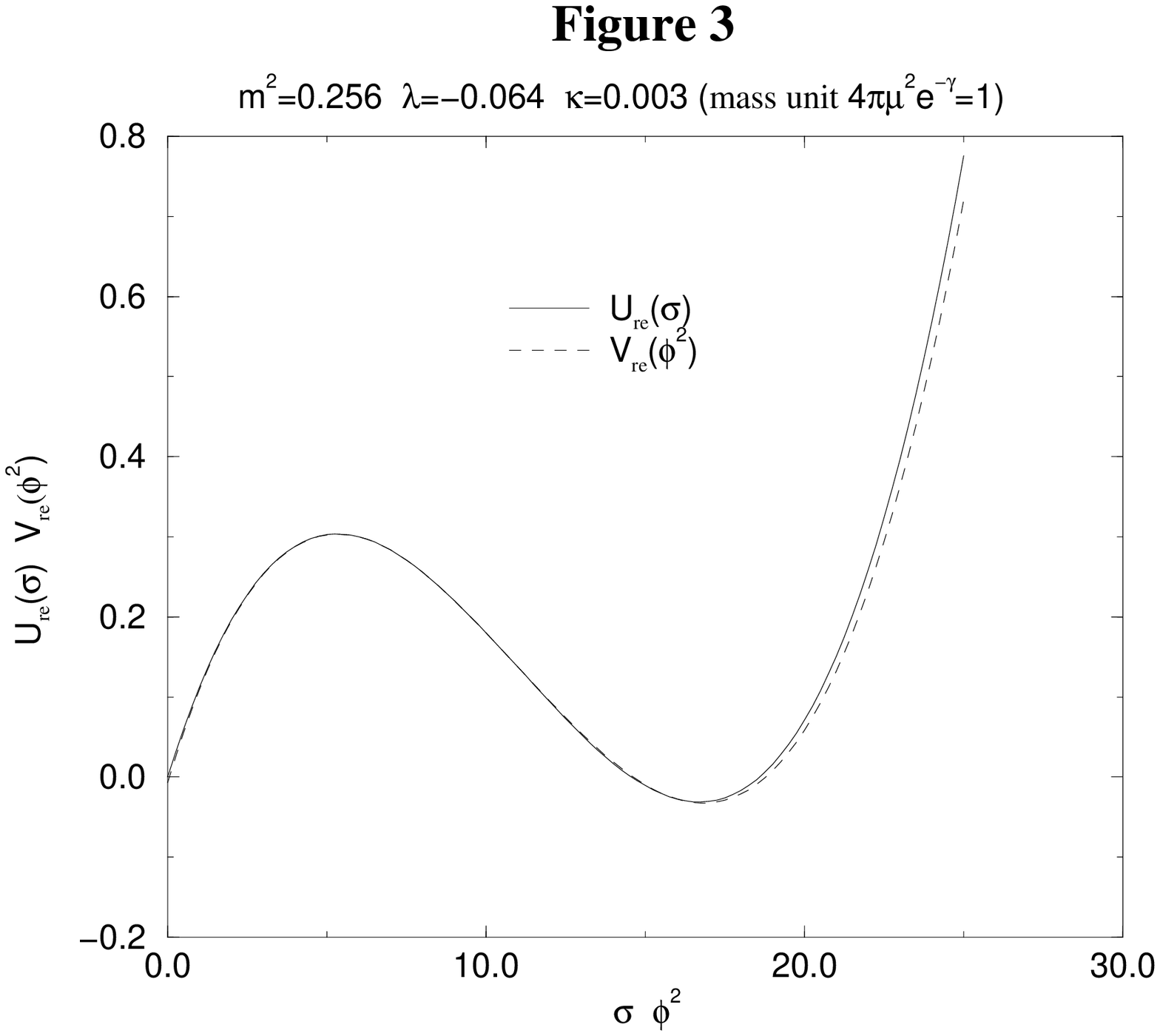}

\end